\documentclass[journal=jacsat, manuscript=article]{achemso}

\usepackage{bm,orcidlink}
\usepackage[version=3]{mhchem}

\author{Jiaji Zhang\,\orcidlink{0000-0003-2978-274X}}
\affiliation{Zhejiang Laboratory, Hangzhou 311100, China}

\author{Jian Liu\,\orcidlink{0000-0002-2906-5858}}
\email{jianliupku@pku.edu.cn}
\affiliation{Beijing National Laboratory for Molecular Sciences,
Institute of Theoretical and Computational Chemistry,
College of Chemistry and Molecular Engineering, Peking
University, Beijing 100871, China}

\author{Lipeng Chen\,\orcidlink{0009-0002-1541-8912}}
\email{chenlp@zhejianglab.com}
\affiliation{Zhejiang Laboratory, Hangzhou 311100, China}

\title{Twin-Space Representation of Classical Mapping Model in the Constraint Phase Space Representation: Numerically Exact Approach to Open Quantum Systems} 

\begin{document}

\begin{abstract}
The \textit{constraint} coordinate-momentum \textit{phase space} (CPS) has recently been developed to study nonadiabatic dynamics in gas-phase and condensed-phase molecular systems.  Although the CPS formulation is exact for describing the discrete (electronic/ vibrational/spin) state degrees of freedom (DOFs), when system-bath models in condense phase are studied, previous works often employ the discretization of environmental bath DOFs, which breaks the time irreversibility and may make it difficult to obtain numerically converged results in the long-time limit.  In this paper, we develop an exact trajectory-based phase space approach by adopting the twin-space (TS) formulation of quantum statistical mechanics, in which the density operator of the reduced system is transformed to the wavefunction of an expanded system with twice the DOFs.  The classical mapping model (CMM) is then used to map the Hamiltonian of the expanded system to its equivalent classical counterpart on CPS.  To demonstrate the applicability of the TS-CMM approach, we compare simulated population dynamics and nonlinear spectra for a few benchmark condensed phase system-bath models with those obtained from the hierarchical equations of motion method, which shows that our approach yields accurate dynamics of open quantum systems. 
\end{abstract}

\maketitle

\section{Introduction}
\label{sec.intro}

Coordinate-momentum phase space formulations of quantum mechanics offer an exact interpretation of quantum systems by mapping quantum operators to continuous-variable functions on phase space. \cite{weyl1927, Wigner1932, Husimi1940, Groenewold1946, Moyal1949, Cohen1966, Lee1995, schroeck1996, Polkovnikov2010, mario2021, Liu2021, He2022, he2024jpcl, Shang2025}
It provides a unified framework to investigate quantum effects by bridging quantum and classical counterpart concepts. 
Beyond conventional coordinate-momentum phase space formulations with infinite boundary for describing systems with continuous variables\cite{weyl1927, Wigner1932, Husimi1940, Groenewold1946, Moyal1949, Cohen1966, Lee1995, Liu2007, Liu2011c, Liu2011b, Liu2011e, Liu2021a, Shao1998}, the generalized coordinate-momentum phase space formulation\cite{liu2016jcp, liu2017, He2019, he2021jpcl, he2021jpca, Liu2021, He2022, Cheng2024, he2024jpcl, Shang2025}, which employs CPS with coordinate-momentum variables for depicting discrete states, has recently been constructed as a rigorous representation for nonadiabatic/composite systems, which involve both continous variables for nuclear DOFs or relatively-low-frequency modes and discrete state DOFs that are not necessarily limited to electronic/vibrational/spin/orbital levels. 
It offers a fundamentally different route from the Schwinger angular momentum theory\cite{Schwinger1965, Sakurai2020} for deriving the renowned Meyer-Miller-Stock-Thoss (MMST) mapping model,{\cite{meyer1979jcp, stock1997prl, sun1998jcp}} which brings the new physical insight that phase space parameter $\gamma_{\rm{ps}}$ of the electronic DOFs (i.e., $F$ discrete state DOFs) of the model ranges from $-1/F$ to $+\infty$ and goes beyond the meaning of the zero-point-energy parameter for the mapping harmonic oscillator originally interpreted by the MMST model\cite{liu2016jcp, He2019, he2021jpcl, he2021jpca}. This generalized coordinate-momentum phase space formulation by Liu and coworkers has been used to develop nonadiabatic field (NaF)\cite{wu2024jpcl, he2024jpcl}, a conceptually new practical approach with independent trajectories for nonadiabatic transition dynamics beyond conventional surface hopping\cite{Tully1990} and Ehrenfest-like dynamics\cite{meyer1979jcp, sun1998jcp, coronado2001cpl, annath2007jcp}, which performs consistently well in both nonadiabatic coupling and asymptotic regions of nonadiabatic transition processes in gas-phase as well as condensed-phase systems. 

When applied to condensed-phase open quantum models, the main difficulty of aforementioned phase space approaches is the dramatically increased number of continuous-variable DOFs due to the environment such as solvent.  To alleviate the computational cost, a commonly used strategy is to treat the environment as a heat bath and the relevant DOFs of interest as a discrete-state system.{\cite{weiss2012, breuer2007}} One then obtains a reduced description of the system (where only discrete state DOFs are included) by tracing out bath DOFs, which is often referred to the quantum master equation (QME) approach.{\cite{tanimura2006jpsj, hagen2009}} The well-known QME approaches include the Redfield theory for the Markovian bath {\cite{schieve2009}} and the numerically exact hierarchical equations of motion (HEOM) for the non-Markovian bath. {\cite{Yan2004,tanimura2020jcp, ye2016wcms, guan2024jcp}} These approaches have been employed to investigate various chemical problems such as the chemical reaction rate,{\cite{zhang2020jcp, shi2011jcp, lindoy2023nc, ke2022jcp}} electron, exciton, or energy transfer, {\cite{zhang2021jcp, shi2009jcp, ishizaki2009pnas, sakamoto2017jpcl, nocera2022jacs, he2021jpcl, he2021jpca, wu2024jpcl}} and photoinduced nonadiabatic transitions.{\cite{ikeda2019jcp, chen2016fd, duan2016jcp, qi2017jcp}}

When the system of the system-bath Hamiltonian model includes both discrete electronic states and only one (or two) continuous nuclear variable(s), it is possible to describe the reduced system by using grids on conventional (Wigner) phase space (of the one or two nuclear variable(s)) for each electronic state.  Tanimura and coworkers have developed the quantum hierarchical Fokker-Planck equation and its multi-state extensions, as the generalization of HEOM for treating open quantum systems.{\cite{tanimura2015jcp, ikeda2019jctc, sakurai2011jpca}} Such an approach has been applied to the investigation of conical intersection,{\cite{ikeda2018cp}} photo-driven Ratchet dynamics,{\cite{iwamoto2019jcp}} and multi-dimensional electronic and vibrational spectroscopy.{\cite{ikeda2017jcp, takahashi2022jcp, takahashi2023jcp}}  However, due to the explicit storage of Wigner phase function and iterative propagation of the kinetic equations, the approach is computationally demanding.  Only when the number of electronic states is small and the number of nuclear DOFs of the reduced system is one or two, it is feasible to apply the approach. 

Several other numerically exact approaches \cite{TDDMRG1,TDDMRG2,MCTDH1,MCTDH2,MCTDH3} treat the open quantum system as a closed system by parametrizing bath DOFs as a set of virtual modes based on some pre-chosen distributions.  These approaches then deal with the effective closed system.  When this technique is applied, the time irreversibility of open quantum systems is not maintained. 
 The technique has been shown to work well for the simulation of short-time dynamics, but the number of bath DOFs should be increased with caution to obtain numerically converged results for long-time dynamics.  When the generalized coordinate-momentum phase space formulation\cite{liu2016jcp, liu2017, He2019, he2021jpcl, he2021jpca, Liu2021, He2022, Cheng2024, he2024jpcl, Shang2025} is applied to benchmark open quantum systems\cite{He2019, he2021jpcl, he2021jpca, Liu2021, He2022, wu2024jpcl, he2024jpcl}, the technique is also implemented.  

The central goal of this work is to develop an exact trajectory-based phase space approach for open quantum systems. To this end, we first employ the twin-space representation of quantum statistical mechanics, where the density operator of the reduced system is transformed to the wavefunction in an expanded system with twice the DOFs.{\cite{borrelli2021wcms, borrelli2019jcp, borrelli2020njp}} In this framework, the dynamical equation for the density matrix is reformulated into a Schr\"{o}dinger-like equation with a complex-valued effective Hamiltonian. We then adopt the classical mapping model (CMM) of the CPS formulation to map the Hamiltonian of the expanded system to its equivalent classical counterpart on quantum phase space with coordinate-momentum variables. The calculation of time-dependent density matrix and multi-time correlation functions follows the general framework of CMM by properly defining operators in the twin-space. Our approach combining twin-space representation and CMM is formally exact without invoking additional approximations, yielding correct long-time dynamics as compared to HEOM. The application in the simulation of population dynamics and nonlinear spectra for a few benchmark condensed phase model systems demonstrates the long-time accuracy of our approach.

\section{Methodology}
\label{sec.theory_open}
Throughout this work, we set the reduced Planck constant $\hbar = 1$ and the Boltzmann constant $k_B =1$.

\subsection{The Liouville space in twin-formulation}
Let us define a double Hilbert space, also referred to as Liouville space, $\mathcal{L} = \mathcal{H} \otimes \tilde{\mathcal{H}}$, where $\mathcal{H}$ is the Hilbert space of a real physical system and $\tilde{\mathcal{H}}$ is the Hilbert space of a fictitious system identical to the original physical system. We introduce the orthonormal basis of $\mathcal{L}$, $\{|k\tilde{l}\rangle\}$, with following relations
\begin{equation}
\langle{k\tilde{l}}|k^{\prime}\tilde{l}^{\prime}\rangle=\delta_{kk^{\prime}}\delta_{\tilde{l}\tilde{l}^{\prime}},\quad\quad\sum_{kl}|k\tilde{l}\rangle\langle{k\tilde{l}}|=1,
\end{equation}
where ${|k\rangle}$ (${|\tilde{k}\rangle}$) is the orthonormal basis of the Hilbert space $\mathcal{H}$ ($\tilde{\mathcal{H}}$). The identity vector $|I\rangle$ is defined as 
\begin{equation}
|I\rangle = \sum_{k} | k \tilde{k} \rangle,
\end{equation} 
which establishes a mapping between the physical space and fictitious tilde space, 
\begin{equation}
\langle{k}|I\rangle=|\tilde{k}\rangle,\quad\quad\langle\tilde{k}|I\rangle=|k\rangle.
\end{equation}

For any operator $A$ acting in the $\mathcal{H}$ space, one can associate a vector in the $\mathcal{L}$ space 
\begin{equation}
|A\rangle=A|I\rangle=\sum_{kl}|k\tilde{l}\rangle\langle{k\tilde{l}}|A|I\rangle=\sum_{kl}\langle{k}|A|l\rangle|k\tilde{l}\rangle=\sum_{kl}A_{kl}|k\tilde{l}\rangle,
\end{equation}
which means that the vector $|A\rangle$ can be represented as a linear combination of basis sets $|k\tilde{l}\rangle$ of $\mathcal{L}$ with coefficients given by $A_{kl}$. For the ease of later derivation, we introduce the projector operator $O_{kl}=|k\rangle\langle{l}|$ in the $\mathcal{H}$ space, which can be represented as a vector in the $\mathcal{L}$ space as 
\begin{equation}
|O_{kl}\rangle=|k\rangle\langle{l}|I\rangle=|k\tilde{l}\rangle.
\end{equation}
We further introduce a state vector $|\rho(t)\rangle=\rho(t)|I\rangle$, where $\rho(t)$ is the density matrix of the system. The expectation value of $A$ can then be defined as the scalar product 
\begin{equation}
\langle{A}\rangle=\langle{A}|\rho(t)\rangle=\langle{I}|A\rho(t)|I\rangle=\mathrm{tr}(A\rho(t))
\end{equation}

Following the formalism of Suzuki\cite{suzuki1991ijmb}, for each Hermitian operator $A$ in the physical space, we define a tilde operator $\tilde{A}$ that is weakly equivalent to $A$ as
\begin{equation}
A |I \rangle = \tilde{A}^{\dagger} | I \rangle ~~
\longrightarrow ~~ A \simeq \tilde{A}^{\dagger}.
\end{equation}
Since $A$ is a Hermitian operator, we have 
\begin{equation}
A \simeq \tilde{A}.
\end{equation}
By employing the tilde conjugation rule, one obtains following relations 
\begin{equation}
\begin{gathered}
(AB)^{\sim} = \tilde{A}\tilde{B} \\
(c_1 A + c_2 B)^{\sim} = c_{1}^{\ast} \tilde{A} + c_2^{\ast} \tilde{B} .
\end{gathered}
\end{equation}
For the commutator of any two operators in the $\mathcal{H}$ space, $[A,B]=AB-BA$, we can reformulate it in the twin-space.  For this purpose, we set $\hat{A}=A-\tilde{A}^{\dagger}$, then 
\begin{equation}
\begin{split}
\hat{A} B |I\rangle &= \left( A - \tilde{A}^{\dagger}\right) B |I\rangle \\  
&= \left( AB - B\tilde{A}^{\dagger}\right) |I\rangle\\
&= \left( AB - BA \right) | I \rangle \\
&=[A,B]|I\rangle,
\end{split}
\label{eq.twin_operator}
\end{equation}
proving the following property
\begin{equation}
[A,B] \simeq \hat{A}B.
\end{equation}
The above equation allows us to rewrite the Liouville equation into a Schr\"{o}dinger-like equation as 
\begin{equation}
\partial_t | \rho(t) \rangle =-i\hat{H}|\rho(t)\rangle=-i(H-\tilde{H})|\rho(t)\rangle.
\end{equation}

\subsection{Reduced density matrix dynamics and its twin-space representation}
We consider a typical system-bath model with the Hamiltonian 
\begin{equation}
H_{tot}=H_s+H_b+H_{sb}.
\end{equation}
The first term $H_s$ is the Hamiltonian of the electronic system,
\begin{equation}
H_{s} = \sum_{j=1}^{F} \varepsilon_{j} |j\rangle \langle j| + 
\sum_{j \ne j^{\prime}} \Delta_{jj^{\prime}}|j\rangle \langle j^{\prime}|,
\label{eq.H_sys_def}
\end{equation}
where $\varepsilon_{j}$ is the energy of the $j$-th state, and $\Delta_{jj^{\prime}}$ is the inter-state coupling. The second term is the Hamiltonian of the heat bath, 
\begin{equation}
H_{b} = \sum_{j=1}^F\sum_{l}\Bigg( \frac{p_{jl}^{2}}{2} + \frac{\omega_{jl}^2 x_{jl}^2}{2}\Bigg),
\end{equation}
where $p_{jl}$, $x_{jl}$, and $\omega_{jl}$ are the dimensionless momentum, coordinate, and frequency of the $l$-th mode of the $j$-th bath. The third term is the system-bath interaction Hamiltonian,
\begin{equation}
H_{sb} =\sum_{j=1}^F V_j \sum_{l} g_{jl} x_{jl},
\end{equation}
with $V_j$ being the system operator, and $g_{jl}$ the coupling strength between the $j$-th state and the $l$-th bath mode which can be specified by the spectral density 
\begin{equation}
J_j(\omega)=\sum_{l}g_{jl}^2\delta(\omega-\omega_{jl}).
\end{equation}
The bath correlation function, which characterizes the effect of the bath on the electronic system, can be defined as 
\begin{equation}
C_j(t) = \int_{0}^{\infty} {\rm{d}}\omega \, J_j(\omega) 
\left[ \coth\left( \frac{\beta\omega}{2} \right) \cos(\omega t) - 
i \sin(\omega t) \right],
\end{equation}
with $\beta=1/T$ being the inverse temperature. In this work, we adopt the Drude spectral density,
\begin{equation}
J_j(\omega) = \frac{2 \lambda_j}{\pi} \frac{\gamma_j \omega}{\gamma_j^2 + \omega^2},
\end{equation}
where $\lambda_j$ is the reorganization energy, and $\gamma_j$ is the inverse correlation time of the $j$-th heat bath, respectively. For the Drude spectral density, one obtains the bath correlation function analytically by using the Pad\'e spectral decomposition,{\cite{jie2011jcp}} 
\begin{equation}
\begin{split}
C_j(t) &= \left[ \lambda_j \gamma_j \cot \left( \frac{\beta\gamma_j}{2} \right) - 
i \lambda_j \gamma_j \right] e^{-\gamma_j t} -
\frac{4\lambda_j \gamma_j}{\beta} \sum_{l=1}^{K}
\frac{\eta_{l} \nu_{l}}{\gamma_j^2 - \nu_{l}^2}
e^{-\nu_l t} \\
&= \sum_{k=0}^{K}c_{jk} e^{-\gamma_{jk} t},
\end{split}
\end{equation}
where $\eta_l$ and $\nu_l$ are the coefficient and frequency of the $l$-th ($l=1,\cdots,K$) Pad\'e term, respectively. We can then derive HEOM that consists of the following set of equations of motion for the auxiliary density operators (ADOs) {\cite{tanimura2020jcp}} 
\begin{equation}
\begin{aligned}
\partial_t \rho_{\bm{n}_1,\cdots,\bm{n}_F}(t) &= -\left[ i H_{s}^{\times} + \sum_{j=1}^F\sum_{k=0}^{K} n_{jk} \gamma_{jk} \right] \rho_{\bm{n}_1,\cdots,\bm{n}_F}(t)-i\sum_{j=1}^F\sum_{k=0}^{K}[V_j,\rho_{\cdots,\bm{n}_j+\bm{e}_{jk},\cdots}(t)]\\
&-i\sum_{j=1}^F\sum_{k=0}^{K}n_{jk}(c_{jk}V_j\rho_{\cdots,\bm{n}_j-\bm{e}_{jk},\cdots}(t)-c_{jk}^{*}\rho_{\cdots,\bm{n}_j-\bm{e}_{jk},\cdots}(t)V_j).
\end{aligned}
\label{eq.heom_def}
\end{equation}
Here, $\bm{e}_{jk}$ is the unit vector along the $jk$-th direction, and we have introduced abbreviation $H_s^{\times}\rho=[H_s,\rho]$. Each ADO is labeled by the index $\bm{n}_j=(n_{j0},\cdots,n_{jK})$, where each element takes a non-negative integer value. The ADO with all indexes equal to zero, $\rho_{\bm{0},\cdots,\bm{0}}$, corresponds to the density operator of the reduced electronic system, while all other ADOs are introduced to take into account non-perturbative and non-Markovian effects. 

The twin-space representation can be applied to open quantum systems by replacing the density operators and super-operators with corresponding wave-functions and operators in the twin-space. Following Ref.~\cite{borrelli2019jcp}, one can derive a set of equations of motion for the auxiliary state vectors in the twin-space analogous to ADOs in Eq.~\eqref{eq.heom_def},
\begin{equation}
\begin{aligned}
\partial_t |\rho_{\bm{n}_1,\cdots,\bm{n}_F}(t)\rangle &= -\left[ i \hat{H}_{s} + \sum_{j=1}^F\sum_{k=0}^{K} n_{jk} \gamma_{jk} \right] |\rho_{\bm{n}_1,\cdots,\bm{n}_F}(t)\rangle \\
&-i\sum_{j=1}^F\sum_{k=0}^{K}(V_j-\tilde{V}_j)|\rho_{\cdots,\bm{n}_j+\bm{e}_{jk},\cdots}(t)\rangle\\
&-i\sum_{j=1}^F\sum_{k=0}^{K}n_{jk}(c_{jk}V_j-c_{jk}^{*}\tilde{V}_j)|\rho_{\cdots,\bm{n}_j-\bm{e}_{jk},\cdots}(t)\rangle.
\end{aligned}
\label{eq.heom_twin_space}
\end{equation}
To further simplify the structure of HEOM, we introduce a set of vectors $|\bm{n}\rangle=|n_{10}n_{11}\cdots{n}_{1K}\\n_{20}\cdots{n}_{FK}\rangle$ and their corresponding bosonlike creation-annihilation operators $b_{jk}^{+}$, $b_{jk}^{-}$ 
\begin{equation}
\begin{aligned}
b_{jk}^{+}|\bm{n}\rangle=\sqrt{n_{jk}+1}|\bm{n}+\bm{e}_{jk}\rangle \\
{b}_{jk}^{-}|\bm{n}\rangle=\sqrt{n_{jk}}|\bm{n}-\bm{e}_{jk}\rangle \\   b_{jk}^{+}b_{jk}^{-}|\bm{n}\rangle=n_{jk}|\bm{n}\rangle,
\end{aligned}
\end{equation}
and the vector 
\begin{equation}
|\Phi(t)\rangle=\sum_{\bm{n}}|\rho_{\bm{n}}(t)\rangle|\bm{n}\rangle,
\end{equation}
Eq.~\eqref{eq.heom_twin_space} can be rewritten in a compact form 
\begin{equation}
\begin{aligned}
\partial_t|\Phi(t)\rangle=&\Bigg[-i\hat{H}_s-\sum_{j=1}^F\sum_{k=0}^K\gamma_{jk}b_{jk}^{+}b_{jk}^{-}-i\sum_{j=1}^F\sum_{k=0}^K\frac{(V_j-\tilde{V}_j)}{\sqrt{n_{jk}+1}}b_{jk}^{+}\\
&-i\sum_{j=1}^F\sum_{k=0}^K \sqrt{n_{jk}}(c_{jk}V_j-c_{jk}^{*}\tilde{V}_j)b_{jk}^{-}\Bigg]|\Phi(t)\rangle \\
=&-i\hat{H}_{\mathrm{eff}}|\Phi(t)\rangle.
\end{aligned}
\label{eq.heom_twin_space_compact}
\end{equation}
$\hat{H}_{\mathrm{eff}}$ can be regarded as the effective Hamiltonian of the expanded system in the twin-space, and $|\rho_{\bm{0}}(t)\rangle=\langle\bm{0}|\Phi(t)\rangle$ 

Within the twin-space representation, the linear response function is defined as{\cite{mukamel2000arpc, mukamel1999}}
\begin{equation}
\begin{split}
R_{1}(t) &= (-i) {\rm{Tr}}\left\{ \mu e^{-i H_{tot}^{\times} t}
\mu^{\times} \rho_{tot}(0) \right\} \\
&= (-i) \, \langle{\bm{0}}|\langle \mu | e^{-i \hat{H}_{\rm{eff}} t} \hat{\mu} | \Phi(0)\rangle,
\end{split}
\label{eq.resp_1st_def}
\end{equation}
where $\rho_{tot}(0)$ is the initial density operator of the total system, $\mu$ is the transition dipole operator, and $\hat{\mu}$ is the twin-space expression of the commutator $\mu^{\times}$. The third-order response function is defined as 
\begin{equation}
\begin{aligned}
R_{3}(t_3, t_2, t_1) &=  (-i)^3  {\rm{Tr}}\left\{ \mu_4 e^{-i H_{tot}^{\times} t_3} \mu_3^{\times} e^{-i H_{tot}^{\times} t_2}
\mu_2^{\times} e^{-i H_{tot}^{\times} t_1} \mu_{1}^{\times} \rho_{tot}(0) \right\} \\
&= (-i)^3 \, \langle{\bm{0}}|\langle \mu_{4} |e^{-i \hat{H}_{\rm{eff}} t_3} \hat{\mu}_{3} e^{-i \hat{H}_{\rm{eff}} t_2} 
 \hat{\mu}_{2}  e^{-i \hat{H}_{\rm{eff}} t_1}  \hat{\mu}_{1} | \Phi(0)\rangle.
\end{aligned}
\label{eq.resp_3rd_def}
\end{equation} 
The rephasing and non-rephasing parts of 2D spectrum are defined by 
\begin{equation}
I_{\rm{R}}(t_2; \omega_3, \omega_1)  = {\rm{Im}} 
\int_{0}^{\infty} {\rm{d}} t_3 \int_{0}^{\infty} {\rm{d}} t_1 \,
 e^{i \omega_3 t_3 - i \omega_1 t_1} \, R_{3}(t_3, t_2, t_1) 
\label{eq.resp_reph_def}
\end{equation}
\begin{equation}
I_{\rm{NR}} (t_2; \omega_3, \omega_1)  = {\rm{Im}} \int_{0}^{\infty} {\rm{d}} t_3 
\int_{0}^{\infty} {\rm{d}} t_1 \, 
e^{i \omega_3 t_3 + i \omega_1 t_1} \, R_{3}(t_3, t_2, t_1) 
\label{eq.resp_nonreph_def}
\end{equation}
with ${\rm{Im}}$ denoting the imaginary part.

There are two main advantages of the twin-space representation of open quantum systems. The first is that the twin-space formalism is exact without invoking additional approximations and universal to any types of QME. The effect of the heat bath is implicitly encoded in the effective Hamiltonian $\hat{H}_{\mathrm{eff}}$, which maintains the time irreversibility of quantum statistical mechanics. The second is that it is a full wave-function based formalism, which facilitates the utilization of wave-function based approaches.

\subsection{Classical mapping model}
\label{sec.theory_naf}
For a multi-state Hamiltonian
\begin{equation}\label{H_electronic}
H=\sum_{n,m=1}^F H_{nm}|n\rangle\langle{m}|,
\end{equation}
the CMM maps the discrete electronic states to coordinates and momentum variables on CPS. The mapped Hamiltonian of Eq.~\eqref{H_electronic} in the phase space reads {\cite{wu2024jpcl, liu2016jcp}}
\begin{equation}
H_{\rm{CMM}}(\bm{x}, \bm{p}) = \sum_{n,m=1}^{F} \left[ \frac{x^{(n)} x^{(m)} + 
p^{(n)} p^{(m)}}{2} - \gamma_{\rm{ps}} \delta_{nm} \right] H_{nm},     
\end{equation}
where $\{\bm{x},\bm{p}\}=\{x^{(1)},\cdots,x^{(F)},p^{(1)},\cdots,p^{(F)}\}$ are the mapping coordinate and momentum variables for the $F$ electronic states, and $\gamma_{\rm{ps}}$ is a phase space parameter defined for CPS. For any system operator $A$, its corresponding phase space function reads 
\begin{equation}
A_{\rm{ps}} (\bm{x}, \bm{p}) = {\rm{Tr}} \left\{ A \, K(\bm{x}, \bm{p}) \right\},
\end{equation}
where the mapping kernel $K(\bm{x},\bm{p})$ is defined as 
\begin{equation}
K(\bm{x}, \bm{p}) =  \sum_{n,m=1}^{F} \left[ \frac{(x^{(n)} + i p^{(n)})
(x^{(m)} - i p^{(m)})}{2} - \gamma_{\rm{ps}} \delta_{nm} \right] 
|n \rangle \langle m| ,
\end{equation}
with $\gamma_{\rm{ps}} \in (-1/F, \infty)$. The time correlation function of two system operators $A$ and $B$ is defined as 
\begin{equation}
\begin{aligned}
C_{AB}(t) &= {\rm{Tr}} \left\{ A(0) B(t) \right\} \\
&= F \int_{\mathcal{S}(\bm{x}, \bm{p})} {\rm{d}}\bm{x}{\rm{d}}\bm{p} \, 
A_{\rm{ps}}(\bm{x}, \bm{p}) \, \bar{B}_{\rm{ps}}(\bm{x}_t, \bm{p}_t),
\end{aligned}
\label{eq.naf_corr_def}
\end{equation}
where $B(t) = \exp(i H^{\dagger}t) B \exp(- i H t)$ is the Heisenberg operator of $B$, and 
\begin{equation}
\bar{B}_{\rm{ps}}(\bm{x}_t, \bm{p}_t) = {\rm{Tr}} \left\{ K^{-1}(\bm{x}, \bm{p}) B(t)\right\},
\end{equation}
with the inverse mapping kernel,
\begin{equation}
K^{-1}(\bm{x}, \bm{p}) =  \sum_{n, m=1}^{F} \left[ \frac{1+F}{(1+F\gamma_{\rm{ps}})^2}\frac{(x^{(n)}+ip^{(n)})(x^{(m)}-ip^{(m)})}{2}   
- \frac{1 - \gamma_{\rm{ps}}}{1 + F \gamma_{\rm{ps}}} \delta_{nm} \right] 
|n\rangle \langle m|. 
\end{equation}
Here, the constraint phase space 
$\mathcal{S}(\bm{x}, \bm{p}):\delta\left( \sum_{n=1}^{F} \left[\frac{(x^{(n)})^2 + (p^{(n)})^2}{2}\right]
- (1+ F \gamma_{\rm{ps}}) \right)$ is defined for the integral on phase space. The equations of motion for phase space variables $(\bm{x}_{t}, \bm{p}_{t})$ read 
\begin{subequations}
\begin{equation}
\partial_t \bm{x}_t = \partial_{\bm{p}_t} H_{\rm{CMM}}(\bm{x}_t, \bm{p}_t) ,
\end{equation}
\begin{equation}
\partial_t \bm{p}_t = - \partial_{\bm{x}_t} H_{\rm{CMM}}(\bm{x}_t, \bm{p}_t) .
\end{equation}
\label{eq.naf_eom_def}
\end{subequations}

The CMM approach can be directly applied to the twin-space representation of the expanded system characterized by the effective Hamiltonian $\hat{H}_{\mathrm{eff}}$ of Eq.~\eqref{eq.heom_twin_space_compact}. To obtain the time evolution of the reduced density matrix $\rho_{mn}(t)=\langle{m}|\rho_{\bm{0}}(t)|n\rangle$, we set the twin-space operators as 
\begin{equation}
\hat{A} = |{\bm{0}} \rangle |\rho_{\bm{0}}(0) \rangle 
\langle \rho_{\bm{0}}(0) | \langle {\bm{0}}| ~~~
\hat{B} = \sum_{j} |{\bm{0}} \rangle | O_{j j} \rangle 
\langle O_{m n}| \langle {\bm{0}}|.
\end{equation}
The time correlation function of two operators $\hat{A}$ and $\hat{B}$ can be written as 
\begin{equation}
\begin{aligned}
C_{AB}(t) &=  {\rm{Tr}} \left\{ \hat{A} e^{i \hat{H}_{\rm{eff}}^{\dagger} t}
\hat{B} e^{-i \hat{H}_{\rm{eff}} t} \right\} \\
&= \sum_{j} {\rm{Tr}} \left\{ |{\bm{0}} \rangle | \rho_{\bm{0}}(0)\rangle 
\langle\rho_{\bm{0}}(0) | \langle {\bm{0}} | e^{i \hat{H}_{\rm{eff}}^{\dagger} t} | 
{\bm{0}} \rangle |O_{j j} \rangle \langle O_{m n}| 
\langle {\bm{0}}| e^{-i \hat{H}_{\rm{eff}} t} \right\} \\
&=\langle{\bm{0}}| \langle O_{mn}| e^{-i \hat{H}_{\rm{eff}} t}  | \rho_{\bm{0}}(0)\rangle |\bm{0}\rangle \sum_{j} \langle\bm{0}|\langle\rho_{\bm{0}}(0)|e^{i\hat{H}_{\mathrm{eff}}^{\dagger}t}|O_{jj}\rangle|\bm{0}\rangle \\
&=\rho_{mn}(t)\sum_j\rho_{jj}(t) \\
&=\rho_{mn}(t),
\end{aligned}
\label{eq.naf_corr_twin}
\end{equation}
where we have used the property of population constraint $\sum_j\rho_{jj}(t)=1$ and the cyclic invariance of trace operation. We can thus evaluate Eq.~\eqref{eq.naf_corr_twin} from Eq.~\eqref{eq.naf_corr_def} following the general procedure of the CMM approach.

The linear response function of Eq.~\eqref{eq.resp_1st_def} can be directly calculated from Eq.~\eqref{eq.naf_corr_twin} by setting 
\begin{equation}
\hat{A} = \hat{\mu} |{\bm{0}}\rangle |\rho_{\bm{0}}(0) \rangle 
\langle \rho_{\bm{0}}(0)| \langle {\bm{0}}| ~~~
\hat{B} = \sum_{j} |{\bm{0}}\rangle| O_{jj} \rangle \langle \mu| \langle {\bm{0}}|.
\end{equation}
To evaluate third-order response function of Eq.~{\eqref{eq.resp_3rd_def}},
we first perform a factorization as
\begin{equation}
R_3(t_3, t_2, t_1) = (-i)^{3} \langle \phi_{de}(t_3) | e^{-i \hat{H}_{\rm{eff}} t_2} | \phi_{ex}(t_1) \rangle, 
\label{eq.naf_resp_3rd}
\end{equation}
where $| \phi_{ex}(t_1) \rangle$ and $
\langle \phi_{de}(t_3) | $ are referred to as excitation and detection states, respectively, 
\begin{equation}
\begin{aligned}
| \phi_{ex}(t_1) \rangle &=  \hat{\mu}_{2} e^{-i \hat{H}_{\rm{eff}} t_1} 
\hat{\mu}_{1} | \rho_{\bm{0}}(0) \rangle |{\bm{0}}\rangle ,\\
\langle \phi_{de}(t_3) |  &=  \langle {\bm{0}}| \langle \mu_{4} | 
e^{-i \hat{H}_{\rm{eff}} t_3} \hat{\mu}_{3} .
\end{aligned}
\label{eq.naf_resp_lr}
\end{equation}
For the excitation state $|\phi_{ex}(t_1) \rangle$, we represent it as a linear combination of complete basis sets of twin space $|k\tilde{l}\rangle$ and hierarchy vector $|\bm{n}\rangle$ as 
\begin{equation}
| \phi_{ex}(t_1) \rangle = \sum_{\bm{n}} \sum_{kl}
\biggl[\langle {\bm{n}} | \langle k\tilde{l}| \phi_{ex}(t_1) \rangle \biggr]
|k\tilde{l}\rangle |\bm{n}\rangle,
\end{equation}
where the coefficients $\langle {\bm{n}} | \langle k\tilde{l}| \phi_{ex}(t_1) \rangle$ can be constructed from Eq.~{\eqref{eq.naf_corr_twin}} by properly assigning operators $\hat{A}$ and $\hat{B}$ as

\begin{equation}
\hat{A} = \hat{\mu}_{1} |{\bm{0}}\rangle |\rho_{\bm{0}}(0)\rangle 
\langle \rho_{\bm{0}}(0)|\langle {\bm{0}}|, \\
~~~ \hat{B} = \sum_{j} |{\bm{0}}\rangle |O_{jj} \rangle \langle
O_{kl}|\langle {\bm{n}}|\hat{\mu}_{2}  .
\end{equation}
This process of constructing coefficients iterates over all the twin-space basis $|k\tilde{l}\rangle$ and hierarchy vector $|\bm{n}\rangle$ at all possible values of $t_1$. It should be noted that the same phase space trajectories can be reused for all the construction process to reduce the computational cost. Following the same procedure, we represent the detection state as   
\begin{equation}
\langle \phi_{de}(t_3) | = \sum_{\bm{n}} \sum_{kl}
\biggl[\langle \phi_{de}(t_3) |k\tilde{l} \rangle |\bm{n}\rangle \biggr]
\langle k \tilde{l} | \langle \bm{n}|, 
\end{equation}
where the coefficients $\langle \phi_{de}(t_3)|k\tilde{l} \rangle |\bm{n}\rangle$ are constructed using following operators   $\hat{A}$ and $\hat{B}$ in Eq.~{\eqref{eq.naf_corr_twin}} 
\begin{equation}
\hat{A} = \hat{\mu}_{3} |{\bm{n}}\rangle |O_{kl}\rangle\langle \rho_{\bm{0}}(0)|
\langle {\bm{0}}| ~~~ \hat{B} = \sum_{j} |{\bm{0}}\rangle |O_{jj} \rangle 
\langle \mu_{4}|\langle {\bm{0}}|.
\end{equation}
Finally, Eq.~{\eqref{eq.naf_resp_3rd}} can be calculated from Eq.~{\eqref{eq.naf_corr_twin}} by setting  
\begin{equation}
\hat{A} = | \phi_{ex}(t_1) \rangle \langle \rho_{\bm{0}}(0)| \langle {\bm{0}}| ~~~
\hat{B} = \sum_{j} |\bm{0}\rangle |O_{jj} \rangle \langle \phi_{de}(t_3) |.
\end{equation}

We note that the definition of $| \phi_{ex} (t_1) \rangle$ and $\langle \phi_{de}(t_3) |$ is similar to that of doorway and window operators in the doorway-window representation of nonlinear response function. {\cite{yan1990pra, yan1989jcp, pios2024jpcl, gelin2021jctc, gelin2011jpcb}} The formulas can be further simplified by employing  
double sided Feynman diagrams of the different Liouville pathways
whenever necessary. {\cite{gelin2010jcp, gelin2022cr}}   

\section{Results and discussion}
\label{sec.result}
We demonstrate the accuracy of the twin-space representation of CMM for a few benchmark condensed phase model systems by comparing results with those from HEOM method. The equations of motion are propagated using the fourth-order Runge-Kutta method (RK4) with a time step of $\delta t$.  Throughout this work, we fix the phase space parameter $\gamma_{\rm{ps}} = 1$ in CMM unless otherwise specified.  We note that any parameter value in the region $(-1/F, \infty)$ leads to the same results because the twin-space representation of CMM in the paper is exact.  In the rest of the work, we use CMM to denote the twin-space representation of CMM for simplicity.  For CMM, all simulations employ $\sim{10^{5}}$ trajectories to get fully converged results.

\subsection{Spin-boson model}
The spin-boson model depicts a two state system linearly coupled to an ensemble of harmonic oscillators; its system Hamiltonian reads 
\begin{equation}
H_s=\frac{\omega_z}{2}\sigma_z+\omega_x\sigma_x.
\end{equation}
The system-bath interaction operator is chosen as $V = \sigma_{x}$. We set $\omega_z=\omega_0$ and $\omega_x=0.5 \omega_0$ using $\omega_0$ as the unit. For the bath parameters, we choose $\lambda = 0.5$, and consider two sets of $\gamma$ and $\beta$: (a) $\gamma = \beta = 3$, and (b) $\gamma = \beta = 6$. Depending on $\gamma$ and $\beta$, the number of Pad\'e frequency terms is varied from 5 to 10, and the number of hierarchies is varied from 5 to 10. The time step is fixed as $\delta t = 0.01$.

\begin{figure}
\centering
\includegraphics[width=0.9\textwidth]{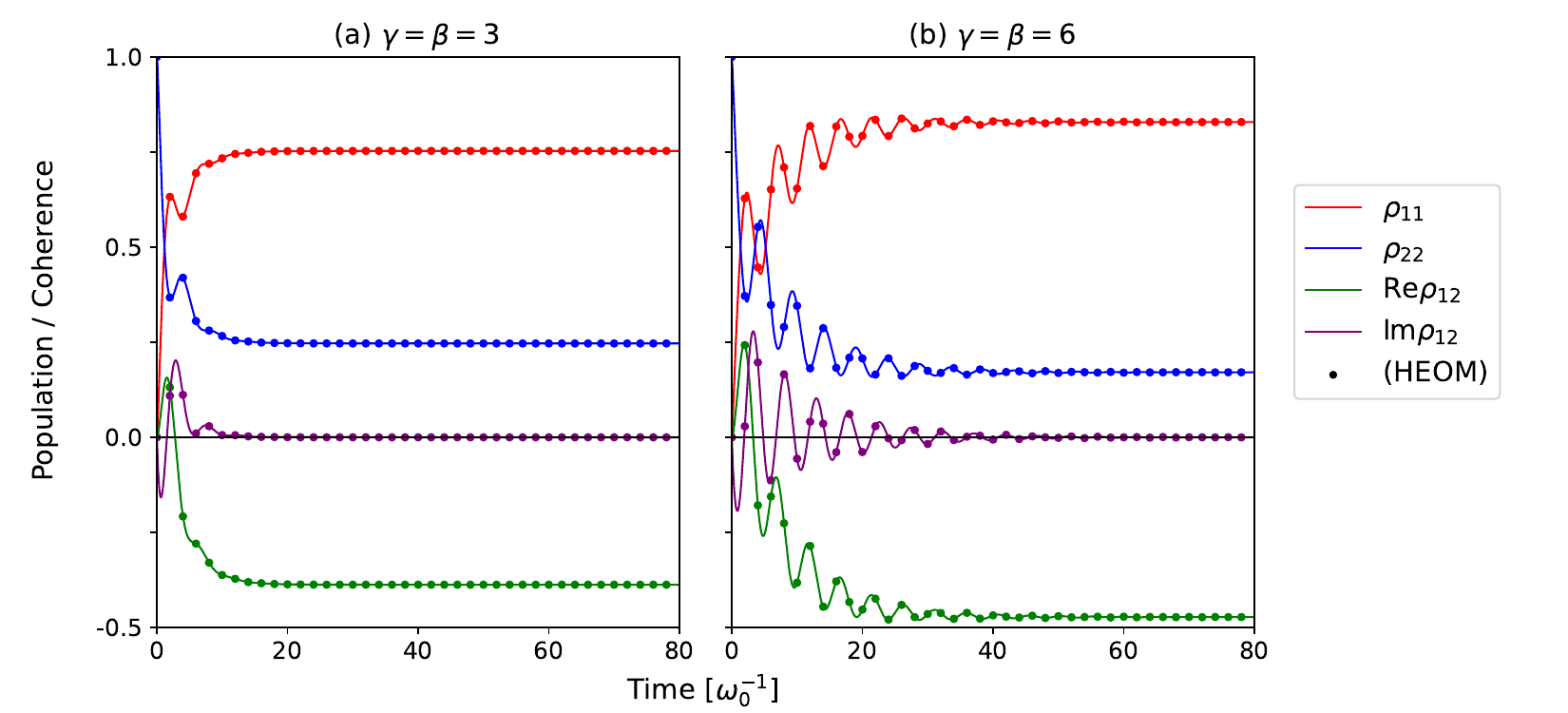}
\caption{Time evolution of $\rho_{nm}(t) = \langle n|\rho_s(t) | m\rangle$ of the spin-boson model calculated by CMM (solid line) and HEOM (dotted line) approaches. Re and Im denote the real and imaginary parts, respectively.}
\label{fig.spin_diag}
\end{figure} 

\begin{figure}
\centering
\includegraphics[width=0.9\textwidth]{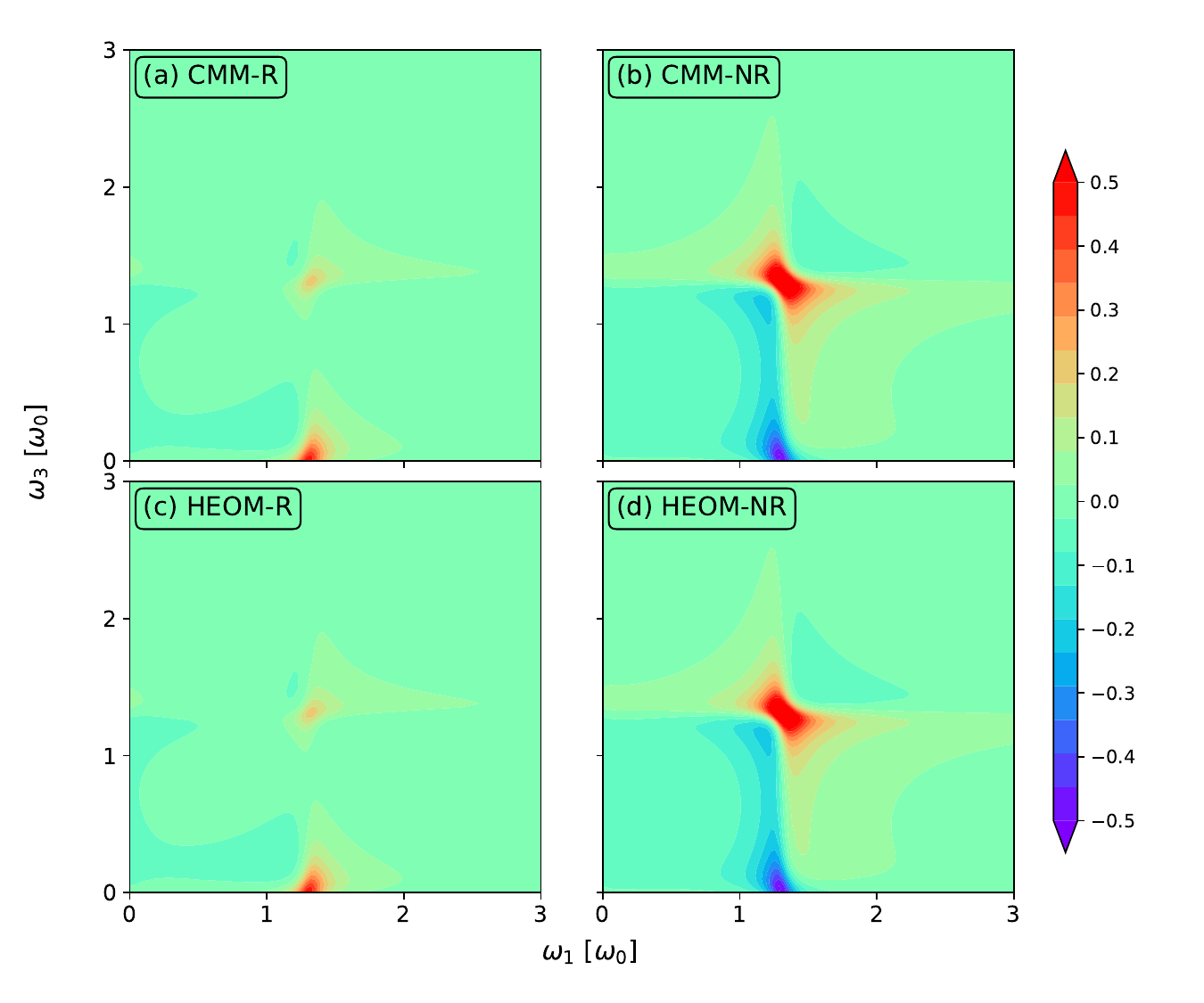}
\caption{Rephasing (R) and non-rephasing (NR) parts of two-dimensional electronic spectra calculated by CMM and HEOM approaches.
The other parameters are the same as Fig.~{\ref{fig.spin_diag}}(b). }
\label{fig.spin_2d}
\end{figure}

Fig.~{\ref{fig.spin_diag}} shows the time evolution of $\rho_{nm}(t) = \langle n|\rho_s(t) | m\rangle$ for the initial state of $\rho_s(0) = |2\rangle \langle 2|$. It is found that our approach yields dynamics in perfect agreement with those from HEOM, illustrating the validity of the twin-space representation of CMM. To further test our approach, we present the rephasing (R) and non-rephasing (NR) parts of two-dimensional electronic spectra at population time $t_2=0$ calculated by CMM and HEOM approaches in Fig.~{\ref{fig.spin_2d}}.
We employ Eqs.~{\eqref{eq.resp_3rd_def}} and {\eqref{eq.naf_resp_3rd}} to calculate the third-order response function via HEOM and CMM approaches, respectively. The transition dipole operator is chosen as $\mu = |1\rangle \langle 2| + |2\rangle \langle 1|$, and other parameters are the same as Fig.~{\ref{fig.spin_diag}} (b). As displayed in Fig.~{\ref{fig.spin_2d}}, our approach reproduces exact spectra.

\subsection{Singlet-fission model}
Singlet-fission is an exciton multiplication process where an excited singlet state generated by irradiation is converted to two triplet excitations. The singlet-fission model that we consider contains 4 electronic states: an electronic ground state $\mathrm{S_0}$, a high-energy singlet state $\mathrm{S_1}$, a charge-transfer state $\mathrm{CT}$, and a multi-exciton state $\mathrm{ME}$. The system Hamiltonian reads
\begin{equation}
H_{s} = \sum_{j = \mathrm{S_0}, \mathrm{S_1}, \mathrm{CT}, \mathrm{ME}} \varepsilon_{j} |j \rangle \langle j|
+ \Delta\left[ |\mathrm{S_1}\rangle \langle \mathrm{CT}| + |\mathrm{CT}\rangle \langle \mathrm{ME}| + {\rm{h.c.}} \right],
\end{equation}
where the site energies $\varepsilon_{j}$ and interstate coupling $\Delta$ are taken from Ref.{\cite{chan2013acr}}, and ${\rm{h.c.}}$ denotes hermitian conjugate. The singlet-fission model has explicitly been tested by HEOM and NaF approaches in Refs. {\cite{he2021jpca, Liu2021, wu2024jpcl, he2024jpcl}}.  The system-bath interaction operator is chosen as $V_{j} = |j\rangle \langle j|$ ($j = \mathrm{S_1}$, $\mathrm{CT}$, and $\mathrm{ME}$), and bath parameters are $\lambda = 0.1$ eV and $\gamma = 0.3$ eV.

\begin{figure}
\centering
\includegraphics[width=0.9\textwidth]{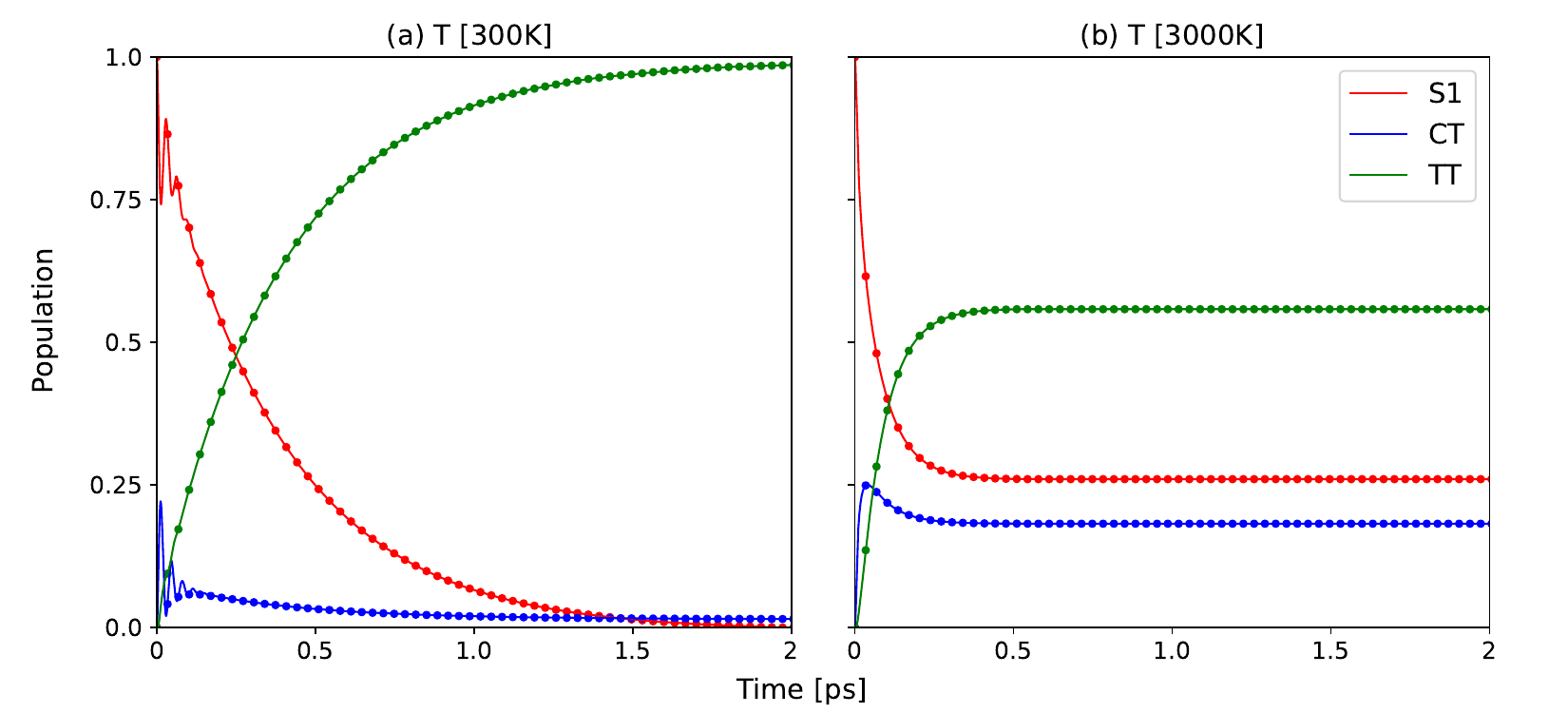}
\caption{Population dynamics of the singlet-fission model at temperatures of (a) 300K and (b) 3000K. Solid and dotted lines correspond to the results obtained from CMM and HEOM approaches, respectively.}
\label{fig.sfm_diag}
\end{figure}

\begin{figure}
\centering
\includegraphics[width=0.9\textwidth]{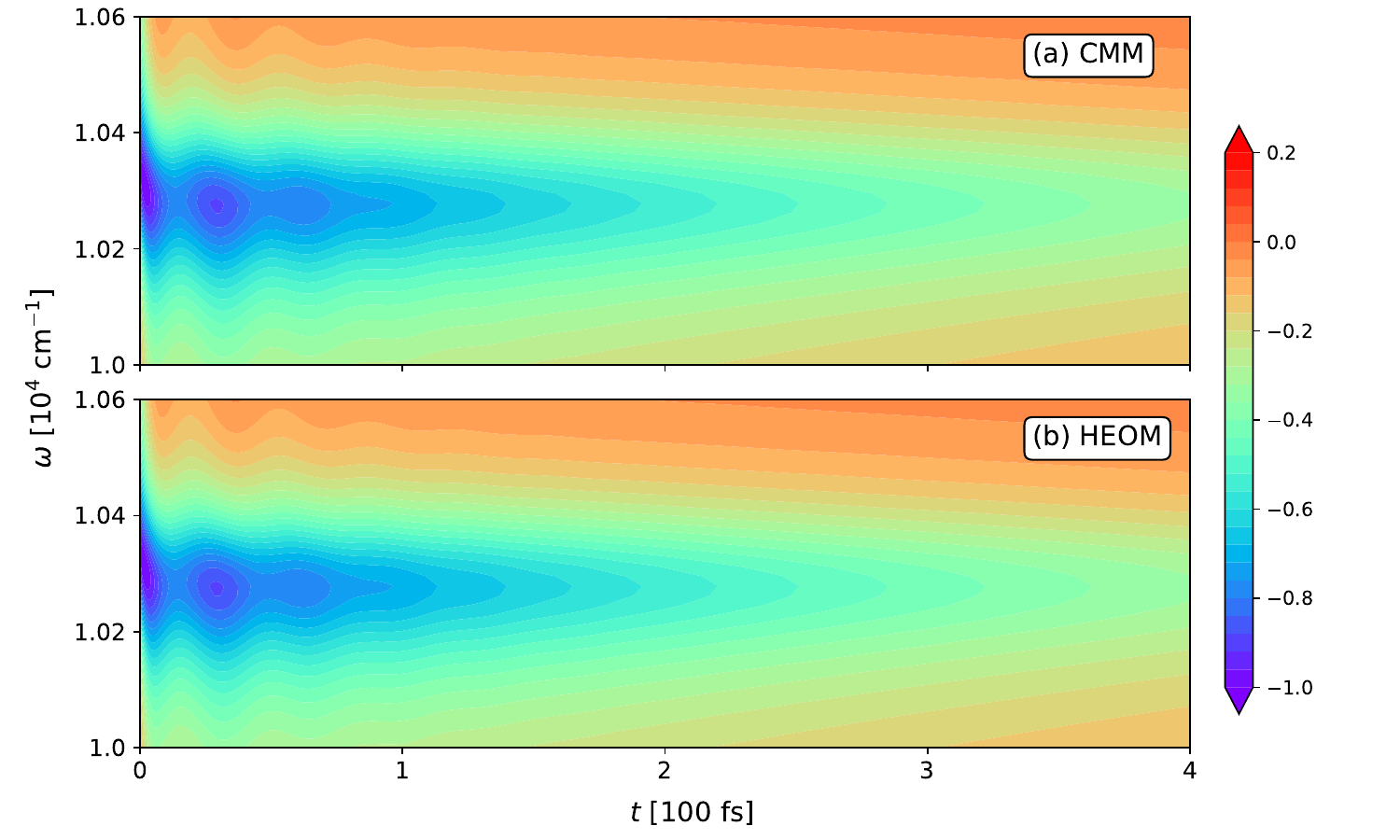}
\caption{TA spectrum of the singlet fission model at 300K calculated from (a) CMM and (b) HEOM approaches, respectively. Other parameters are the same as Fig.~{\ref{fig.sfm_diag}}(a). }
\label{fig.sfm_300}
\end{figure}

\begin{figure}
\centering
\includegraphics[width=0.9\textwidth]{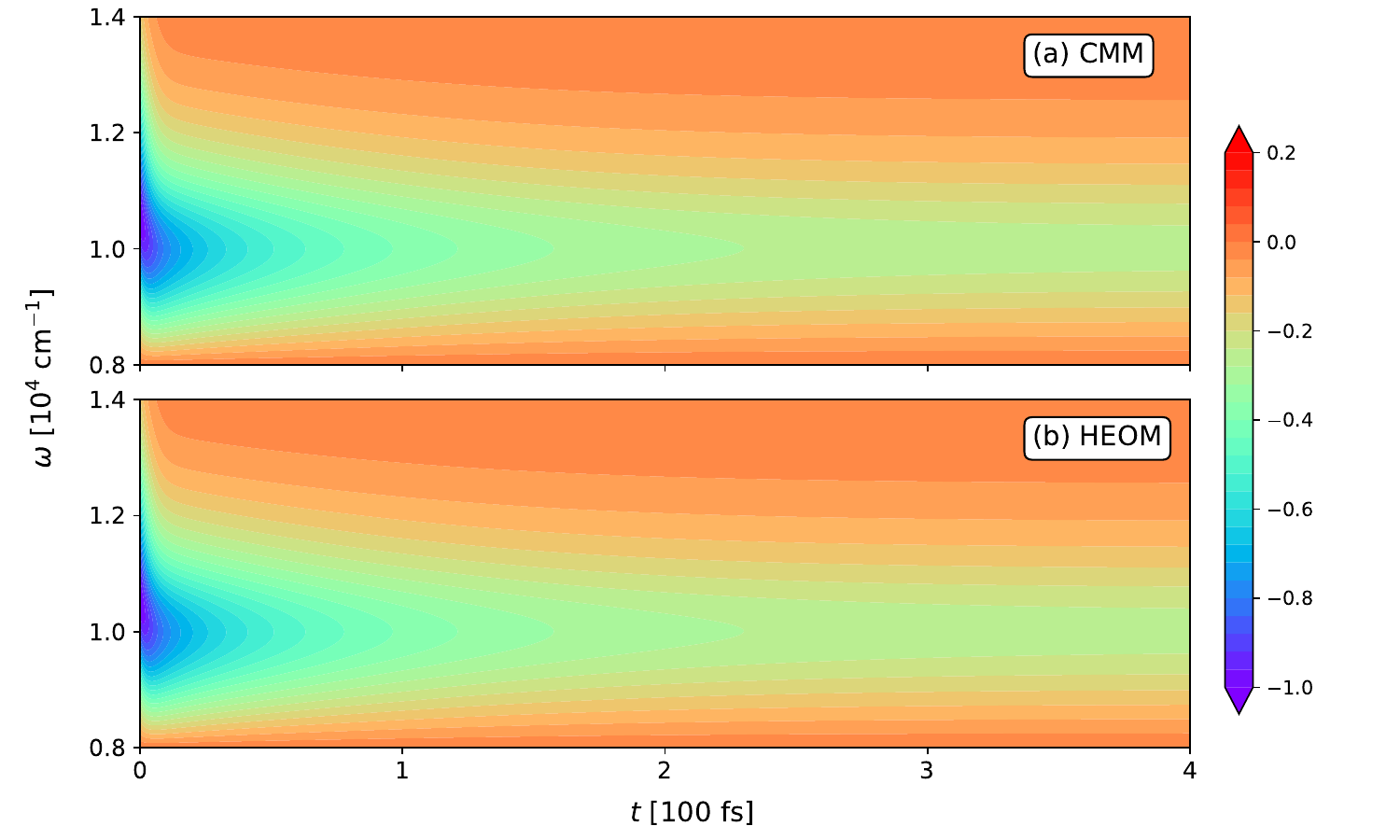}
\caption{TA spectrum of the singlet fission model at 3000K calculated from (a) CMM and (b) HEOM approaches, respectively.
Other parameters are the same as Fig.~{\ref{fig.sfm_diag}}(b). }
\label{fig.sfm_3000}
\end{figure}

Fig.~{\ref{fig.sfm_diag}} shows population dynamics of the singlet fission model at temperatures of (a) 300K and (b) 3000K, using $\rho_{s}(0) = |\mathrm{S_1}\rangle \langle \mathrm{S_1}|$ as the initial state. The CMM approach is capable of capturing the correct short-time as well as long-time dynamics. One can clearly observe that the oscillation between $\mathrm{S_1}$ and $\mathrm{CT}$ states within the first 30 fs disappears as the temperature increases from 300K to 3000K. We also calculate the transient absorption (TA) spectrum, which can be obtained from the third-order response function as  
\begin{equation}
I_{\rm{TA}}(t, \omega)  = {\rm{Im}} \int_{0}^{\infty} {\rm{d}}t_3 \, e^{-i \omega t_3} \, R_{3}(t_3, t, 0).
\end{equation}
In the singlet fission model, since only $\mathrm{S_1}$ state is optically bright, the transition dipole operator can be chosen as $\mu=|\mathrm{S_0}\rangle\langle\mathrm{S_1}|+|\mathrm{S_1}\rangle\langle\mathrm{S_0}|$. The TA spectra at 300K and 3000K are plotted in Figs.~{\ref{fig.sfm_300}} and {\ref{fig.sfm_3000}}, respectively. In both figures, the spectrum reflects the population dynamics as  shown in Fig.~{\ref{fig.sfm_diag}}. The spectrum calculated from the CMM method matches perfectly with that from the HEOM approach.

\subsection{Seven-site model for the Fenna-Matthews-Olson monomer}

\begin{figure}
\centering
\includegraphics[width=0.9\textwidth]{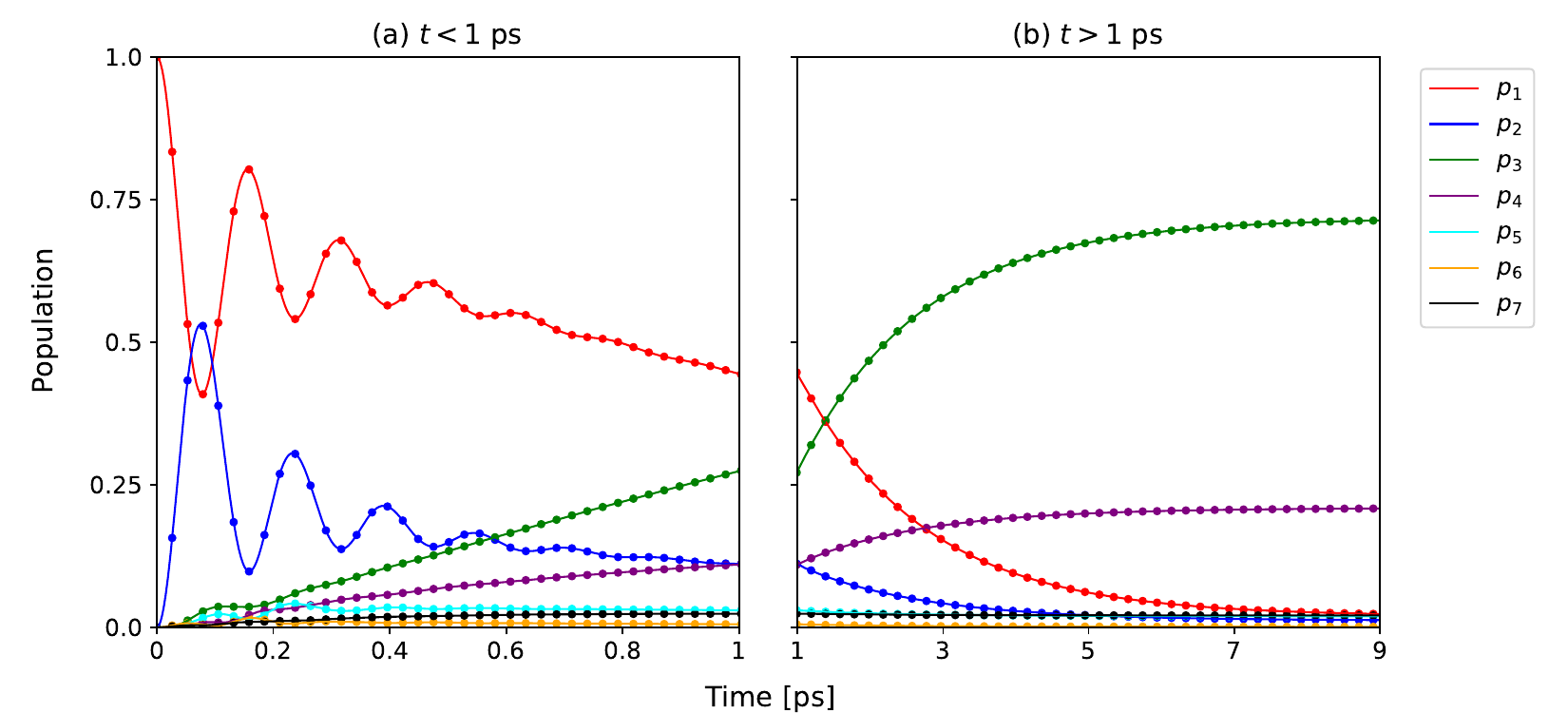}
\caption{Population dynamics of the FMO model at 77K calculated from CMM (solid lines) and HEOM (dotted lines) approaches, respectively. Panel (a) corresponds to the short time dynamics up to 1ps, and panel (b) corresponds to the long time dynamics up to 9ps.}
\label{fig.fmo_diag}
\end{figure} 

\begin{figure}
\centering
\includegraphics[width=0.9\textwidth]{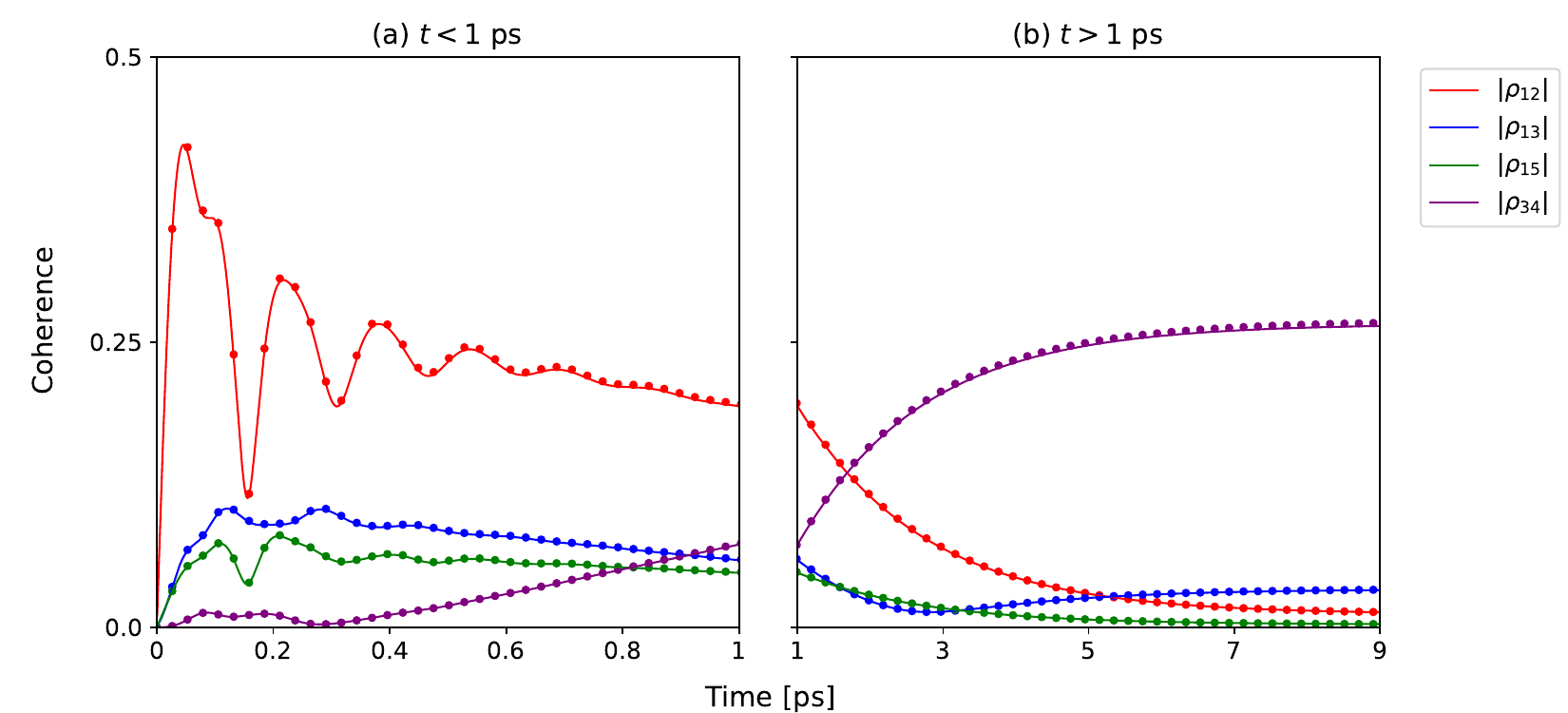}
\caption{Coherence dynamics of the FMO model at 77K calculated from CMM (solid lines) and HEOM (dotted lines) approaches, respectively. 
All the settings are the same as Fig.~{\ref{fig.fmo_diag}}.}
\label{fig.fmo_offdiag}
\end{figure}

The Fenna-Matthews-Olson (FMO) protein complex is a prototype system to study the excitonic energy transfer process in photosynthetic organisms. We consider the exciton model of apo-FMO complex;{\cite{ishizaki2009pnas}} its system Hamiltonian can be written as
\begin{equation}
H_{s} = \sum_{j=1}^{F} \varepsilon_{j} |j\rangle \langle j| + 
\sum_{j \ne j^{\prime}} \Delta_{j j^{\prime}}|j\rangle \langle j^{\prime}|,
\label{eq.H_fmo_1ex}
\end{equation}
with $\varepsilon_{j}$ being the excitation energy of the $j$th site and $\Delta_{jj'}$ the interstate couplings. Each site is coupled to an individual bath, i.e., $V_j=|j\rangle\langle{j}|$. The bath parameters are $\lambda_{j} = 35$ ${\rm{cm}}^{-1}$, $\gamma_j = 100$ ${\rm{cm}}^{-1}$, and temperature $T = 77$ K. The number of Pad\'e frequency terms is 1, and HEOM is truncated at the hierarchy level of 10. In Fig.~{\ref{fig.fmo_diag}}, we plot the population dynamics up to 9ps for the initial state of  $\rho_{s}(0) = |1\rangle \langle 1|$, demonstrating that the CMM approach reproduces exact short time dynamics as well as long-time steady state.  
The corresponding coherence dynamics is presented in Fig.~{\ref{fig.fmo_offdiag}}.

\begin{figure}
\centering
\includegraphics[width=0.9\textwidth]{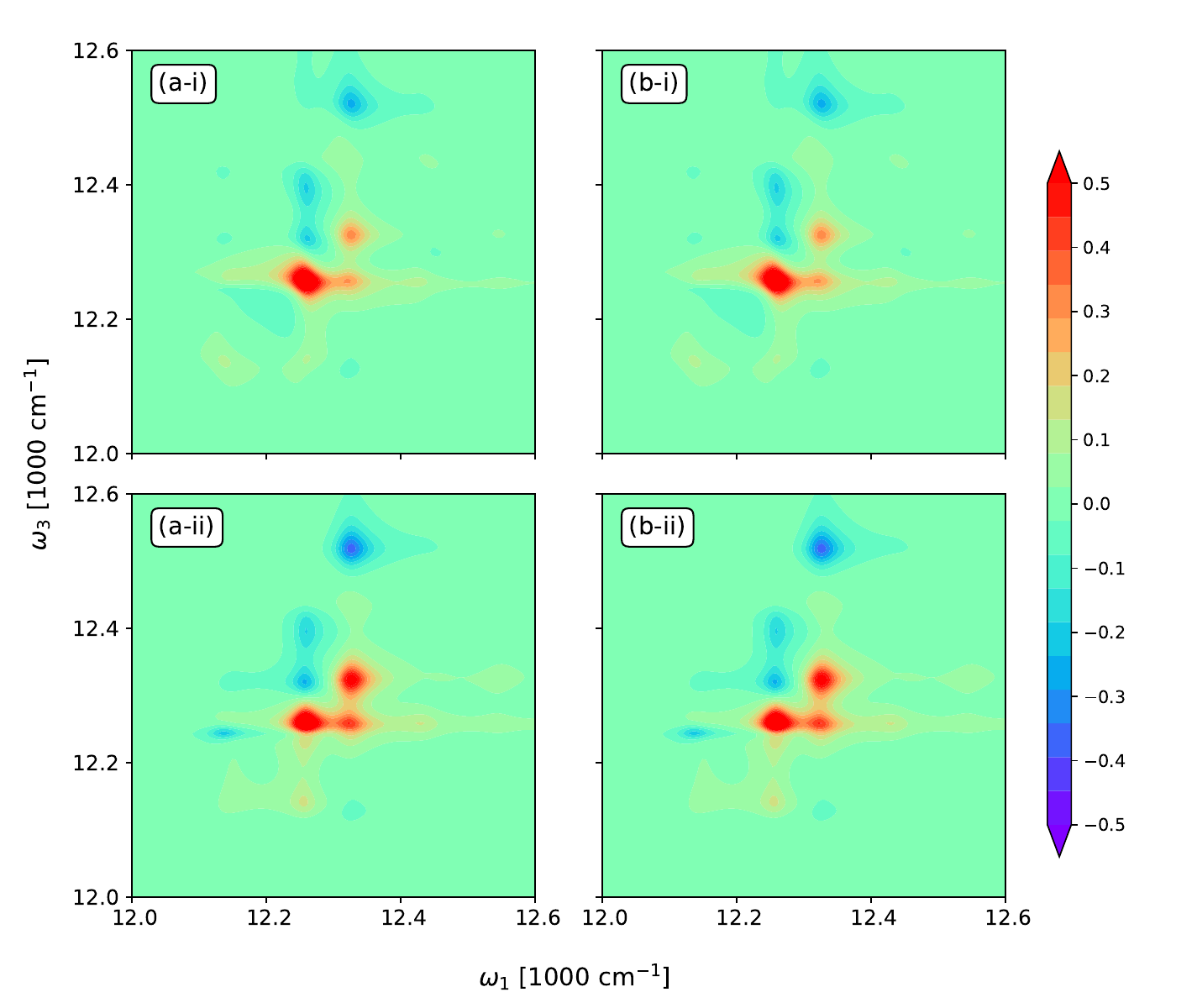}
\caption{2DES of the FMO model at population times of (i) $t_2 = 0$ and (ii) $t_2 = 160 ~ {\rm{fs}}$. Panels (a) and (b) correspond to the results obtained from CMM and HEOM approaches. In all figures, the peaks are normalized with respect to the maximum intensity of figure (a-i) without applying the arcsinh scaling.}
\label{fig.fmo_2des}
\end{figure} 

We also calculate two-dimensional electronic spectrum (2DES) to reveal the excitation energy relaxation process of the FMO complex. To simulate 2DES, one needs to include the two-exciton states $(|jk\rangle)$, which represent the simultaneous excitation at two sites, into the system Hamiltonian, i.e., 
\begin{equation}
H_{s} = \varepsilon_{g} |g\rangle \langle g| + H_{\mathrm{1ex}} + H_{\mathrm{ 2ex}}.
\end{equation}
Here, $|g\rangle$ denotes the ground state, $H_{\mathrm{1ex}}$ is the one-exciton Hamiltonian as defined in Eq.~{\eqref{eq.H_fmo_1ex}}. The two-exciton Hamiltonian $H_{\mathrm{2ex}}$ can be constructed by considering following relations {\cite{hein2012njp, chen2011jcp}}
\begin{equation}
\langle jk | H_{\mathrm{2ex}} | jk\rangle = \langle j | H_{\mathrm{1ex}} |j \rangle + \langle k | H_{\mathrm{1ex}} |k \rangle,
\end{equation}
\begin{equation}
\begin{aligned}
\langle ij | H_{\mathrm{2ex}} | kl\rangle &= \delta_{ik}(1 - \delta_{jl}) \langle j | H_{\mathrm{1ex}} | l \rangle + \delta_{il}(1 - \delta_{jk}) \langle j | H_{\mathrm{1ex}} | k \rangle \\
&+\delta_{jk}(1 - \delta_{il}) \langle i | H_{\mathrm{1ex}} | l \rangle + \delta_{jl}(1 - \delta_{ik}) \langle j | H_{\mathrm{1ex}} | k \rangle,
\end{aligned}
\end{equation}
where $\delta_{jk}$ denotes the Kronecker delta function. The transition dipole operator is the sum of one-exciton and two-exciton contributions
\begin{equation}
\mu = \mu_{\mathrm{1ex}} + \mu_{\mathrm{2ex}}.
\end{equation}
Here, $\mu_{\mathrm{1ex}}$ is the one-exciton contribution as defined by 
\begin{equation}
\mu_{\mathrm{1ex}} = \sum_{j} \mu_{j} |j \rangle \langle g| + {\rm{h.c.}},
\end{equation}
where $\mu_{j} = \vec{d}_{j} \cdot \vec{l}$, with $\vec{d}_{j}$ being the  transition dipole moment of the $j$th site, and 
$\vec{l}$ the polarization of laser pulse, respectively. The two-exciton contribution $\mu_{\mathrm{2ex}}$ can be constructed as follows
\begin{equation}
\langle ij | \mu_{\mathrm{2ex}}| k\rangle  = \delta_{ik} \langle j | \mu_{\mathrm{1ex}} | g\rangle 
+ \delta_{jk} \langle i | \mu_{\mathrm{1ex}} | g\rangle .
\end{equation}
The total 2DES can be obtained as the summation of rephasing and non-rephasing contributions
\begin{equation}
I_{\rm{2DES}}(t_2; \omega_3, \omega_1) = I_{\rm{R}}(t_2; \omega_3, \omega_1)  + 
I_{\rm{NR}}(t_2; \omega_3, \omega_1) .
\label{eq.resp_corr_def}
\end{equation}

In the simulation of 2DES, we consider a specific polarization $\vec{l} = (1, 0, 0)$. We use one Pad\'e frequency term and truncate HEOM at the hierarchy level of 4, which is enough to obtain converged spectra. In Fig.~{\ref{fig.fmo_2des}}, we present 2DES at population times of $t_2 = 0$ and $ t_2 = 160 ~ {\rm{fs}}$ obtained from CMM and HEOM approaches. Here, positive peaks correspond to  contributions from the ground state bleach and the stimulated emission, while negative peaks represent the contribution from the excited state absorption. The CMM approach yields 2DES in perfect agreement with those from the reference HEOM.

\subsection{Quantum morse oscillator model}
In this section, we consider a dissipative quantum morse oscillator model where a quantum morse oscillator is coupled to a dissipative heat bath; its system Hamiltonian reads  {\cite{ishizaki2006jcp}}
\begin{equation}
H_{s} = \frac{P^2}{2}  + D \, \left( 1 - e^{-\alpha Q} \right)^2,
\end{equation}
where $P$ and $Q$ are the dimensionless momentum and coordinate, $D$ is the dissociation energy, and $\alpha$ represents the curvature of the potential. The system parameters are chosen as $D = 81600$ ${\rm{cm}}^{-1}$ and $\alpha = 0.1$. We use the lowest 6 eigenstates of the morse oscillator for the following simulation. The system-bath interaction operator is chosen as $V = Q$. For the simulation of two-dimensional infrared spectrum (2DIR), we consider two sets of bath parameters which exhibit different lineshapes.{\cite{cho2019, jasen2019jcp}} One is the spectral diffusion regime with $\lambda = 30$ ${\rm{cm}}^{-1}$ and $\gamma = 8$ ${\rm{cm}}^{-1}$, representing the slow bath dynamics; the other is the motional narrowing regime with $\lambda = 160$ ${\rm{cm}}^{-1}$ and $\gamma = 800$ ${\rm{cm}}^{-1}$, representing the fast bath dynamics. The simulation of 2DIR follows the same procedure of 2DES, except that the dipole operator is defined as  $\mu=Q$.     

\begin{figure}
\centering
\includegraphics[width=0.9\textwidth]{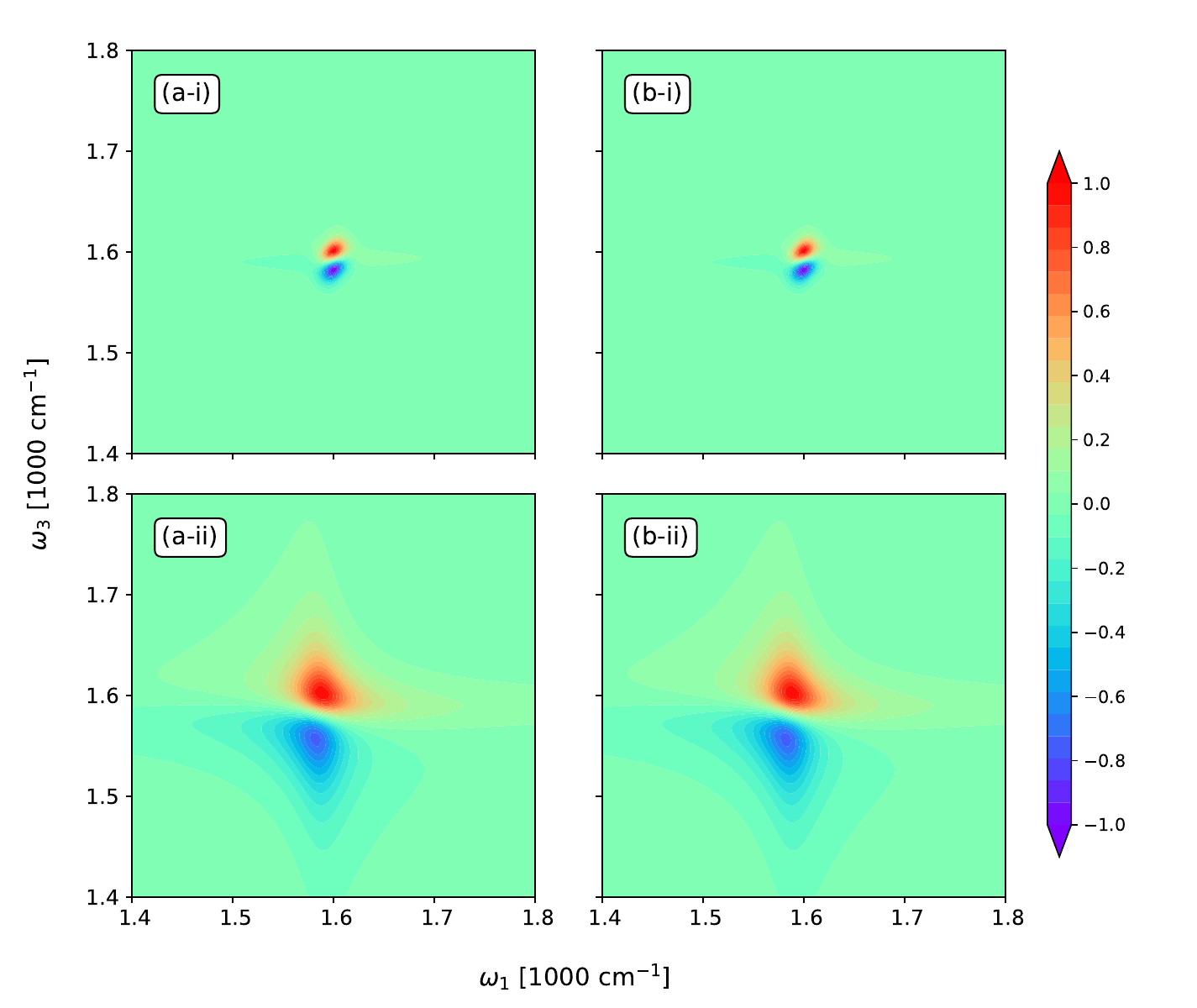}
\caption{2DIR of the quantum morse oscillator model at temperature T=300K for (i) spectral diffusion regime and (ii) motional narrowing regime. Panels (a) and (b) correspond to the results obtained from CMM and HEOM approaches. The peaks calculated by CMM and HEOM approaches are normalized with respect to the maximum intensity of Figs.~(a-i) and (b-i), respectively.}
\label{fig.morse_2dir}
\end{figure} 

In Fig.~{\ref{fig.morse_2dir}}, we plot the 2DIR at temperature T=300K for both spectral diffusion and motional narrowing regimes as calculated from CMM and HEOM approaches. As shown in Fig.~{\ref{fig.morse_2dir}}, the node line between positive and negative peaks is parallel to the diagonal line for the spectral diffusion regime, while it becomes almost horizontal for the motional narrowing regime. The CMM approach reproduces the 2DIR calculated by the HEOM method.

\subsection{Vibronically coupled dimer model}

The vibronically coupled dimer model has been widely used to simulate the two-dimensional electronic vibrational  spectroscopy (2DEVS).{\cite{pallavi2019jpcl}} 
In a 2DEVS experiment, the system interacts with two UV-vis pulses followed by a IR pulse. 
The resultant spectrum monitors the evolution of correlation between electronic and vibrational DOFs. We consider an exciton dimer, with an IR mode local to each of the monomers (labeled A and B, respectively). The system Hamiltonian is defined as  
\begin{equation}
H_{s} = \sum_{j=0,1,2} (\varepsilon_{j} + U_{j}) |j\rangle \langle j | 
+ \sum_{j=1,2}\sum_{j \ne k} \Delta_{j k} |j\rangle \langle k |,
\end{equation}
with
\begin{equation}
U_{j} = \sum_{X = A, B}\left\{ \frac{p_{j,X}^2}{2} + \frac{\omega_{j,X}}{2}
\left( q_{j, X} - q_{j, X}^0 \right)^2\right\}.
\end{equation}
Here, we consider three electronic states $|j\rangle$ ($j=0,1,2$): an electronic ground state $|0\rangle$ where both A and B are in their respective ground states, and two excited states $|1\rangle$ and $|2\rangle$ representing that only A or B is electronically excited. $\varepsilon_j$ and $\Delta_{jk}$ are the state energy and interstate electronic couplings. $p_{j, X}$, $q_{j, X}$, $\omega_{j, X}$ and $q_{j, X}^0$  are the dimensional momentum, coordinate, frequency, and displacement of the vibrational mode $X$ $(X = A, B)$ for the $j$th electronic state, respectively.  
The system-bath interaction operator is chosen as $V_{j} = q_{j, A} + q_{j, B}$. 

In our simulation, we represent $H_s$ in the site basis. We consider only lowest two vibrational state for each mode and denote basis sets as  $|j_{g_A, g_B} \rangle$, $|j_{e_A, g_B} \rangle$, and $|j_{g_A, e_B} \rangle$ ($j=0,1,2$), where $g_X$ and $e_X$ are the ground and first excited vibrational states of mode $X$. We further approximate the site basis as a product form,  $|j_{\alpha_A, \beta_B} \rangle = |j\rangle \otimes |\alpha_A^{j}\rangle \otimes |\beta_B^{j}\rangle$, where $|\alpha_X^j\rangle$ is the $\alpha$th $(\alpha = g, e)$ eigenstate of mode $X$ for the $j$th electronic state. The overlap of vibrational states for different electronic states is described by the Huang-Rhys factor $\sigma$, i.e.,
\begin{equation}
\begin{gathered}
\langle g_X^{0} | g_X^{j} \rangle = e^{-\sigma / 2} \\
\langle g_X^{0} | e_X^{j} \rangle = \sqrt{\sigma}e^{-\sigma / 2} \\
\langle g_X^{j} | e_X^{0} \rangle = -\sqrt{\sigma} e^{-\sigma / 2} \\
\langle e_X^{0} | e_X^{j} \rangle = (1 - \sigma) e^{-\sigma / 2}
\end{gathered}
\end{equation}
for $X = A,B$ and $j = 1,2$. The numerical values of system parameters are taken from Ref.~{\cite{pallavi2019jpcl}}, with 
$\varepsilon_{0} = 0$, $\varepsilon_{1} = 12000$ ${\rm{cm}}^{-1}$, 
$\varepsilon_{2} = 12900$ ${\rm{cm}}^{-1}$, $\Delta_{12} = 250$ ${\rm{cm}}^{-1}$, 
$\omega_{0, X} = 1030$ ${\rm{cm}}^{-1}$, 
$\omega_{1, X} = \omega_{2, X} = 950$ ${\rm{cm}}^{-1}$, and $\sigma = 0.0025$. The bath parameters are chosen as $\lambda_j = 30$ ${\rm{cm}}^{-1}$, $\gamma_{j} = 100$${\rm{cm}}^{-1}$ ($j=1,2$), and temperature $T = 77$ K.

\begin{figure}
\centering
\includegraphics[width=0.9\textwidth]{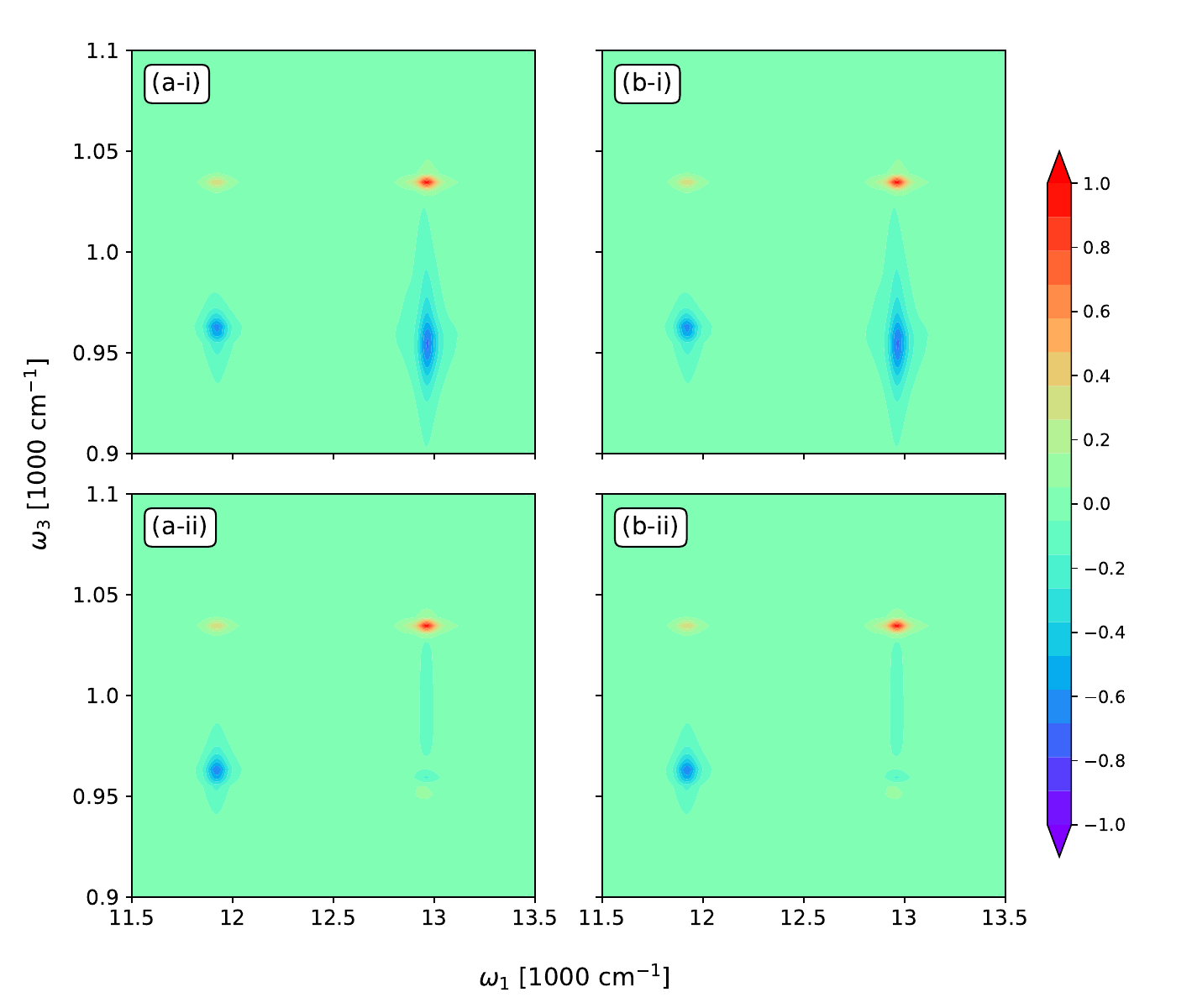}
\caption{2DEV spectrum of a vibronically coupled dimer model at population times of (i) $t_2 = 0$, and (ii) $t_2 = 550$ fs calculated by (a) CMM and (b) HEOM approaches, respectively. In all figures, the peaks are normalized with respect to the maximum intensity of figure (a-i). }
\label{fig.dimer_2dev}
\end{figure}

The 2DEV spectrum is simulated using the transition dipole operators, $\mu_1 = \mu_2 = \mu_{ele}$, and $\mu_{3} = \mu_{4} = \mu_{vib}$, which are defined as 
\begin{subequations}
\begin{equation}
\mu_{ele} = \sum_{j = 1, 2}\left( |0\rangle \langle j| + |j\rangle \langle 0| \right),
\end{equation}
\begin{equation}
\mu_{vib} = \sum_{j = g, 1, 2} q_{j, A} + q_{j, B} .
\end{equation}
\end{subequations}
In Fig.~{\ref{fig.dimer_2dev}}, we show 2DEV spectrum at population times of (i) $t_2 = 0$ and (ii) $t_2 = 550$ fs calculated from (a) CMM and (b) HEOM approaches, respectively. 
In the $\omega_1$ axis, the frequencies at $11900$ ${\rm{cm}}^{-1}$ and $12950$ ${\rm{cm}}^{-1}$ correspond to the electronic transitions  $|0\rangle \to |1\rangle$ and $|0\rangle \to |2\rangle$, respectively. 
In the $\omega_3$ axis, the frequencies at $1040$ ${\rm{cm}}^{-1}$ and $960$ ${\rm{cm}}^{-1}$ correspond to vibrational transitions of the electronic ground state $|g_X^{0}\rangle \to |e_X^{0}\rangle$ and the electronic excited states $|g_X^{j}\rangle \to |e_X^{j}\rangle$ $(j = 1,2)$, respectively. 
The time evolution of  cross peaks reflect the energy relaxation process. As shown in panels (a) and (b) of Fig.~{\ref{fig.dimer_2dev}}, the CMM results are in perfect agreement with those from the HEOM method.


\section{Conclusion}
\label{sec.conclusion}

In summary, we have developed a new trajectory-based phase space approach for the open quantum system by integrating the twin-space representation of quantum statistical mechanics into the framework of the classical mapping model in the CPS formulation. The proposed method employs the twin-space formalism to transform the density operator of the reduced system into a form that can be treated as the wavefunction of an expanded system with twice of the discrete-state DOFs. The classical mapping model is then adopted to map the Hamiltonian
of the expanded system to its equivalent classical counterpart on CPS.  The strengths of our theory have been demonstrated by simulating population dynamics and nonlinear spectra of a few benchmark condensed phase model systems. It is shown that our approach combining the twin-space representation and classical mapping model yields correct long-time dynamics, rendering it a promising tool for studying open quantum systems.  

This work lays the foundation for further exploration of trajectory-based approaches in the generalized coordinate-momentum phase space formulation.  For instance, when the reduced system involves both discrete-state variables and continuous variables, our current approach can be generalized to a more practical approximate approach on quantum phase space by integrating out the bath DOFs.  Recent development of advanced laser technologies allows the manipulation of specific quantum pathways to magnify desired spectroscopic features in various physical and chemical processes. It is of great interest to extend our approach to scenarios involving time-dependent Hamiltonians, which find broad applications in the quantum control and strong-field spectroscopy.{\cite{gelin2010jcp, norambuena2024prl}} The theoretical framework is beyond the scope of the response function theory, one thus needs to treat the system-field interaction nonperturbatively. Work in these directions is in progress.

\section*{Acknowledgments}
J.Z. and L.P.C. acknowledge the support from the National Natural Science Foundation of China (No. 22473101).  J.L. thanks the support from the National Science Fund for Distinguished Young Scholars Grant No. 22225304.


\section*{Data availability}
The data that support the findings of this study are available from 
the corresponding author upon reasonable request.

%

\bibliography{ref_list}

\end{document}